\documentclass[11pt]{article}
\usepackage{amsmath}
\usepackage{color}
\usepackage{amssymb}
\usepackage{graphicx,psfrag,epsf}
\usepackage{enumerate}
\usepackage{url,appendix} % not crucial - just used below for the URL 
\usepackage{longtable}
\usepackage{framed}
\usepackage{rotating}
\usepackage{caption}
\usepackage{pbox}
\usepackage{booktabs}

\definecolor{marro}{rgb}{0.6,0,0}

\newcommand{\Prob}{\mathbb{P}}

\newcommand{\bmath}{\boldsymbol}

\newcommand*{\MyIndent}{\hspace*{0.5cm}}
\newcommand{\rowgroup}[1]{\hspace{0em}\textbf{#1}}

\begin{document}
\bibliographystyle{plain}
\title{\textbf{Regularization and Hierarchical Prior Distributions for Adjustment with Health Care Claims Data: Rethinking Comorbidity Scores}}

\author{Jacob V Spertus$^{1}$, Samrachana Adhikari$^{1}$, and Sharon-Lise T Normand$^{1,2}$ \\
	\small{1: Department of Health Care Policy, Harvard Medical School}\\ \small{2: Department of Biostatistics, Harvard TH Chan School of Public Health}}
\date{}
\maketitle

% Running headers of paper:
\markboth%
% First field is the short list of authors
{J. V. Spertus, S. Adhikari, and S-L. T. Normand}
% Second field is the short title of the paper
{Bayesian Regularization and Hierarchical Adjustment}

\begin{abstract}
	Health care claims data refer to information generated from interactions within health systems. They have been used in health services research for decades to assess effectiveness of interventions, determine the quality of medical care, predict disease prognosis, and monitor population health. While claims data are relatively cheap and ubiquitous, they are high-dimensional, sparse, and noisy, typically requiring dimension reduction. In health services research, the most common data reduction strategy involves use of a comorbidity index --  a single number summary reflecting overall patient health. We discuss Bayesian regularization strategies and a novel hierarchical prior distribution as better options for dimension reduction in claims data. The specifications are designed to work with a large number of codes while controlling variance by shrinking coefficients towards zero or towards a group-level mean. A comparison of drug-eluting to bare-metal coronary stents illustrates approaches. In our application, regularization and a hierarchical prior improved over comorbidity scores in terms of prediction and causal inference, as evidenced by out-of-sample fit and the ability to meet falsifiability endpoints.
\end{abstract}

\maketitle
\section{Introduction}
\label{sec:intro}

\indent

Health services researchers focus on assessing the effectiveness of treatments and policy interventions, characterizing the quality of medical care delivered to specific populations, and developing payment systems, with the goal of improving health outcomes. A fundamental data source used in health services research is administrative billing claims data. Claims data refer to a wide range of administrative databases capturing health care utilization information for reimbursement purposes and are generated by routine interactions between patients, the health system, and insurance providers \cite{iezzoni1997}. This type of data stands in contrast to other formats, like clinical registry data, which are prospectively collected for the explicit purpose of later analysis and are of considerably higher quality \cite{massdac}\cite{sarrazin2012}\cite{iezzoni1997}. However claims data have many desirable features: they are generated during the normal work flow and thus avoid additional effort and expense; they provide an accounting of denominator populations to ensure rates are well-defined; and they have standard definitions to help with consistency of the information collected across the nation and around the world.  

In this article, we focus on claims data comprised of alpha-numeric codes for conditions from the International Classification of Diseases (9th Clinical Modification, [ICD-9]) system \cite{icd}. The structure of ICD-9 codes is a tree-like hierarchy with each code constructed as a binary variable. The codes contain 3-5 alpha-numeric strings where the first 3 characters define a group of diagnoses (or procedures) and the additional two characters provide further specificity. The ICD-9 codes include approximately 17,000 distinct codes at level 5, many of which represent rare conditions that are unlikely to be observed in any finite population. Coding practices can also vary across health delivery settings such that codes are noisy representations of true conditions. A revision of ICD-9 codes (ICD-10) took effect in October 2015 in the United States, and contains approximately 68,000 codes having a similar hierarchical structure to the ICD-9 system. Thus, while claims data are relatively cheap and ubiquitous, they are high-dimensional, sparse, and noisy. Consequently, some form of dimension reduction for inference is required.
 
Researchers have proposed a variety of dimension reduction approaches when using claims data. The most common strategies either use clinical expertise to select a small set of codes and include separate binary indicators for each discrete diagnosis or procedure, or implement standard variable selection approaches such as step-wise regression. Because these approaches may omit important variables, inclusion of a summary measure, called a comorbidity index or score, helps with adjusting for other health features \cite{sharabiani2012}. Two of the most frequently used comorbidity indices include the Charlson and the Elixhauser \cite{charlson1987}\cite{elixhauser1998}. These indices were constructed by selecting and validating ICD-9 codes based on a combination of clinical expertise and statistical analysis. The Elixhauser index has been shown to predict long-term clinical outcomes better than the Charlson index \cite{sharabiani2012}. Both indices may be implemented as a set of dummy variables indicating the presence of groups of conditions (e.g. hypertension, diabetes, etc) if feasible or as a single continuous variable constructed from a weighted sum of the individual condition groups \cite{charlson1987}\cite{walraven2009}. A recent paper by Gilbert et al argued that using a comorbidity index with fixed historical weights is as good as including individual claims coefficients in causal survival analysis \cite{gilbert2017}. However the assumptions associated with the strategy are rather strong, inferences may be compromised by ignoring uncertainty from the estimation of the weights, and the findings do not translate to other approaches like propensity score analyses or logistic regression for mortality prediction.

Our point of departure from typical approaches is to consider the dimension reduction task as one of regularization. Fundamentally some outside information, typically an assumption that some of the conditions have little or no effect on the outcome of interest, must be incorporated when the number of variables available exceeds the capacity of the data for appropriately precise estimation.

 We note that other sophisticated methods have been developed to use high-dimensional claims data for various analysis goals. Syed et al proposed using a support vector machine to enhance risk prediction using current procedural terminology codes \cite{syed2011}. Perotte et al proposed exploiting the hierarchical nature of ICD-9 codes using a support vector machine for diagnosis code assignment \cite{perotte2014}. Singh used the group lasso to improve outcome prediction by accounting for the ICD-9 hierarchy in a frequentist setting \cite{sing2015}. In the causal inference domain, Schneeweiss et al developed a frequentist algorithm to select claims confounders based on their prevalence and univariate associations with both treatment and outcome \cite{schneeweiss2009}.

%\subsection{Drug-Eluting vs Bare-Metal Stents}  \label{sec:des_bms}
Our work is motivated by determining the comparative effectiveness of coronary stents once released into the U.S. market. Stents are small scaffolds placed into the heart via a percutaneous coronary intervention (PCI) in order to treat arteries blocked by plaque and keep them clear and open. Drug-eluting stents (DESs) are coated in a drug designed to keep plaque from re-forming, while bare-metal stents (BMSs) have no such coating. Clinical trials have consistently shown that DESs are efficacious as they reduce the need for target vessel revascularization (TVR), i.e. follow-up procedures in the same artery, but have no benefit on survival \cite{bonaa2016}\cite{stettler2007}. However, DESs necessitate adherence to prolonged dual-antiplatelet therapies to prevent rare but fatal stent thrombosis. Several patient characteristics such as age, comorbidity, and socio-economic factors may impact adherence to dual-antiplatelet therapy, and thus such patients are likely to be implanted with BMSs \cite{spertus2006}. Because these characteristics also tend to increase the risk of adverse outcomes, observed differences in average outcomes between BMS and DES treated patient groups are heavily confounded. Even after statistical adjustments various observational analyses have generally found that DES provides a survival benefit in usual care populations \cite{mauri_circ2008}\cite{tu2007}. Although there has historically been speculation that this benefit was real but not observed in randomized trial populations, a recent quasi-randomization analysis and the results of a large pragmatic clinical trial make a strong case that any observed survival benefit is due to residual confounding \cite{venkitachalam2011}\cite{bonaa2016}. As a result, it remains an open challenge for observational causal inference to accurately estimate the effects of DES versus BMS using observational data. It may be possible to achieve better results by including claims data in some form.

While we utilize a Bayesian propensity score framework, the ideas apply to other techniques such as frequentist propensity scores or Bayesian outcome modeling. A Bayesian approach is particularly well aligned with high-dimensional inference with claims data, as it provides a number of intuitive tools for model building and incorporates all major sources of uncertainty, which may ultimately lead to better inferential performance \cite{spertus2018}. Our focus is on improving over a typical analysis using comorbidity indices and to this end we make three interdependent arguments: (1) administrative claims data can help analysts meet the ignorability assumption in observational studies, but standard comorbidity indices may be insufficient; (2) using a considerably larger and more granular set of claims data may lead to better results compared with a comorbidity index; and (3) regularization techniques are very useful for controlling variance when using a high-dimensional set of claims data. Combining these principles, we find that employing high-dimensional claims data with Bayesian regularization techniques leads to significant improvement over a comorbidity index in terms of treatment model fit and causal inference. 

In section \ref{sec:causal_inference} we briefly remind the reader the necessary assumptions for causal inference with observational data and review propensity score approaches. Our proposed methods are presented in detail in section \ref{sec:proposed_approaches}. Section \ref{sec:massdac_analysis} illustrates various strategies for dimension reduction by revisiting the DES vs BMS problem. We conclude with a discussion of the advantages of regularization over comorbidity indices for causal inference, as well as potential improvements and modifications for future research.

\section{Causal Inference and Propensity Scores}
\label{sec:causal_inference}

\subsection{Causal Inference}
 
Causal inference involves various assumptions, three of which include stable unit treatment value assumption (SUTVA), positivity, and treatment assignment ignorability. We focus on binary treatments where a new treatment is compared to a standard treatment. In this setting, SUTVA implies that treatment does not vary within treatment groups and that the treatment received by one subject does not affect the outcomes of another. Positivity requires that every subject has a chance of receiving the new treatment and we assume that this probability is bounded away from 0 and 1. Positivity can be empirically tested in that blatant violations are apparent by examination or modeling of the data. Propensity scores help facilitate assessing the positivity assumption \cite{petersen2010}. Finally, ignorability assumes that potential outcomes are independent of treatment assignment \cite{rosenbaum1983}. In observational designs, investigators must adjust for confounders, variables that affect both treatment assignment and outcome, to uncover the causal effect of a particular treatment. This task is among the most challenging in observational causal inference and has inspired many attempts to define methods, either statistical or expertise driven, to select confounders for adjustment \cite{brookhart2006}\cite{schneeweiss2009}\cite{spertus2018}. High-dimensional datasets may provide important variables that help investigators meet the ignorability assumption. In particular, health care claims data provide an exciting source of potential confounders, presenting a host of opportunities as well as challenges for researchers.

\subsection{Propensity Scores}
\label{sec:propensity_scores}
 
Propensity scores are the probabilities of receiving the new treatment relative to the standard treatment. They are a commonly used tool for observational causal inference.  One particularly important feature involves facilitating an {\em outcome free analysis} in that they separate outcome modeling from treatment modeling which can help to maintain objectivity \cite{rubin2008}. This feature is often a requirement by regulators, such as the U.S. Food and Drug Administration, who place a premium on objective findings \cite{levenson2012}. Propensity scores can be particularly useful in high-dimensional settings when event rates are low or when the number of confounders is large relative to number of events. As a result, many investigators interested in observational causal inference with high-dimensional data have focused on how to build propensity score models. Variable selection is the standard response to dealing with high-dimensional data, and researchers have historically focused on selecting important confounders based on subject matter expertise and rules of thumb \cite{brookhart2006}\cite{stuart2013}. More algorithmic variable selection techniques have also been explored, including the high-dimensional propensity score algorithm of Schneeweiss et al, but this uses outcome information to select variables \cite{schneeweiss2009}. Other approaches have proposed more technical variance reduction strategies through machine learning methods, such as elastic net regularization, tree-based algorithms, and learning ensembles \cite{spertus2016}\cite{mccaffrey2004}\cite{vanderlaan2011}. Each of these methods has shown promise in estimating causal effects from large databases. We use a Bayesian framework because it propagates uncertainty from the propensity score model and provides intuitive and sophisticated variance reduction strategies via prior distributions, an advantage that we leverage through a hierarchical prior specification.  
 
 \subsubsection{Bayesian Propensity Scores}
 \label{sec:bps}
 
We adopt a Bayesian treatment model and maintain an outcome free analysis, although jointly modeling the propensity score and outcomes has been proposed \cite{mccandless2009}\cite{saarela2015}. The particular algorithm we use resembles frequentist propensity score weighting \cite{spertus2018}. The algorithm uses a conjugate prior-likelihood pair in the form of a beta-binomial: if $Y$ represents the total number of subjects having an event out of $n$ subjects, e.g., $Y \stackrel{\mbox{\footnotesize iid}}{\sim} \mbox{Bin}(n, p),$ then assuming $p$ arises from $\mbox{Beta}(a,b)$ \textit{a-priori}, yields \textit{a-posteriori} $p \mid {Y} \sim \mbox{Beta}(a + Y, b + n - Y)$.  We use this fact to integrate over draws from the propensity score model while conditionally drawing from the outcome model, so that uncertainty is incorporated from both the propensity score model and the outcome model. 

Briefly, let $Y_i$ denote binary outcome for subject $i$, $X_i$ binary treatment, and $\pi_i$ the propensity score,  $\Prob(X_i = 1)$. The parameter $\bmath \pi$ is the vector of propensity scores for all subjects, $\bmath X$ is the vector of treatment indicators, $\bmath Y$ the vector of outcomes, and ${\bf D} = \{\bmath{X}, \bmath{Y}\}$ their concatenation into an $n \times 2$ matrix. Together, $\bmath \pi$ and ${\bf D}$ are sufficient to estimate an unbiased treatment effect under an ignorable treatment assignment mechanism. Let $p_1$ and $p_0$ be the marginal probability of the event in treatment group 1 and 0 respectively. These can be estimated in an unbiased fashion by weighting the original populations according to the inverse of their propensity scores.

We let $\widetilde{Y}(1)$ denote the total number of potential events in treatment group 1 (all subjects for which $X_i = 1$), and $\widetilde{Y}(0)$ the number of events in treatment group 0. By assuming a binomial likelihood for $\widetilde{Y}(1)$ and $\widetilde{Y}(0)$, and selecting flat priors $p_0$ and $p_1 \sim \mbox{Beta}(1,1)$, we can take advantage of conjugacy to get closed form solutions for the posterior distributions for $p_0$ and $p_1$. The priors are augmented by weighted counts of successes and failures in each treatment group to generate beta posteriors , $p_g \mid {\bf D}, {\bmath \pi} \sim \mbox{Beta}  \big(a^*_g, b^*_g \big)$ for $g = 0, 1$ where 
\begin{align*}
a^*_g &= 1 +  \gamma_g \bigg(g\sum_{i=1}^n \frac{X_i Y_i}{\pi_i} + (1-g)\sum_{i=1}^n \frac{(1-X_i) Y_i}{1-\pi_i}\bigg),\\ b^*_g &= 1 + \gamma_g \bigg(g\sum_{i=1}^n  \frac{X_i (1-Y_i)}{\pi_i} + (1-g)\sum_{i=1}^n \frac{(1-X_i) (1-Y_i)}{1-\pi_i}\bigg)
\end{align*}
and $a^*_g~(b^*_g)$ describes the weighted number of events (non-events) in treatment group $g$. The quantity $\gamma_g$ is used to normalize the weighted population to its original size and is defined as $\gamma_g = g \sum_{i=1}^n \frac{(1-X_i) (1-Y_i)}{1-\pi_i} + (1-g)\sum_{i=1}^n \frac{(1-X_i) (1-Y_i)}{1-\pi_i}$. The causal parameter, such as the average treatment effect, can be calculated as simple functions of draws from the respective posteriors of $p_0$ and $p_1$ - for example the average treatment effect defined by the posterior risk difference, $\Delta = p_1 - p_0$. Estimates can be obtained using Markov Chain Monte Carlo simulation and incorporate the uncertainty from both the propensity score and outcome models (see in Spertus et al \cite{spertus2018}). 

\section{Proposed Risk Adjustment Approaches}
\label{sec:proposed_approaches}

We now consider estimating propensity scores $\pi_i = \Prob(X_i | \boldsymbol{B}_i, \boldsymbol{C}_i)$, where $\boldsymbol{B}_i$ is a vector of baseline confounders, for example demographic or clinical covariates from a clinical registry, and $\boldsymbol{C}_i$ is a vector of 4-digit ICD-9 diagnosis codes. We use 4-digit rather than 5-digit codes for simplicity and without loss of generality.  We define the propensity score model as

\begin{align}
\pi_i &= \mbox{logit}^{-1} \big\{\beta_0 + \boldsymbol{\beta}_B \boldsymbol{B}_i + \boldsymbol{\beta}_C \boldsymbol{C}_i \big\}. \label{eqn:ps_model}
\end{align}

The intercept $\beta_0$ and coefficients associated with the baseline confounders $\boldsymbol{\beta}_B$ are standard logistic regression parameters. We set a non-informative prior for the intercept as $\boldsymbol{\beta}_0 \sim \mathcal{N}(0, 10)$. The 95\% interval for $\beta_0$ is about [-20,20], which on the logit scale encompasses the vast majority of possible values for the marginal treatment rate. We assume that each component of $\boldsymbol{\beta}_B$ is distributed \textit{a priori} as $t_5(0, 2.5)$ and components are mutually independent. This prior is referred to as ``weakly informative'' in the Bayesian data analysis literature \cite{gelman2014}. It places the majority of prior mass within [-5,5] though heavy tails allow larger estimates if the data warrants. Weakly informative priors help produce a well-defined posterior but otherwise do little shrinkage compared to a maximum likelihood estimate. Having specified priors for the intercept and baseline variables, we turn to various approaches to handling the claims variables $\boldsymbol{C}_i$ using regularization.

\subsection{Regularization via Prior Distributions}
\label{sec:claims_regularization}

\subsubsection{Weakly Informative Prior Distributions}
A natural choice for the prior distribution of $\boldsymbol{\beta}_C$ is weakly informative such that $\boldsymbol{\beta}_C \sim t_5(0, 2.5)$. These priors could be applied to groups of codes at the 3-digit level, or to more granular 4-digit codes. The latter is likely to have higher variance but less bias, and we explore both possibilities in our application. However, alternatives may be required because it is likely that the majority of the ICD-9 codes are rare and have no relationship with the treatment assignment, so they will introduce considerable variance. 

\subsubsection{Sparsity-Inducing Prior Distributions}

An alternative {\em sparsity-inducing} prior is the horseshoe prior where for the $j$th component of $\boldsymbol{\beta}_C$. We assume the associated coefficient is
\begin{align}
\beta_{Cj} &\sim \mathcal{N}(0, \lambda_j^2 \tau^2); ~\mbox{where} ~\lambda_j \sim \mbox{Cauchy}^+(0,1)~\mbox{and}~\tau \sim \mbox{Cauchy}^+(0,1). 
\end{align}
The notation $\mbox{Cauchy}^+(0,1)$ denotes a half-Cauchy distribution defined over the positive real line. The horseshoe performs global shrinkage through $\tau$ and local shrinkage through $\lambda_j$, allowing signals to remain near their maximum likelihood estimates, while shrinking noise variables aggressively towards zero. It has shown impressive predictive properties in simulations and applied problems \cite{carvahlo2010}\cite{peltola2014}. For easier sampling, we replace the priors for $\lambda_j$ with $t_3^+(0, 1)$ priors which users find have little effect on coefficient estimates in practice \cite{piironen2015}. 

Under this specification, 50\% of the prior mass on $\beta_{Cj}$ is placed between -.4 and .4, which is quite close to zero. However the 95\% prior interval is about (-11,11), reflecting the fact that the horseshoe places a lot of prior weight near zero but also has ``fat tails" in order to avoid shrinking signals.

\subsubsection{Hierarchical Prior Distributions}
\label{sec:hierarchical_model}

As an alternative to standard weakly informative or sparsity-inducing priors, we propose a hierarchical prior on $\boldsymbol{\beta}_C$ that takes advantage of the natural tree-like hierarchy in ICD-9 codes. Assuming $p$ elements for $\boldsymbol{C}_i$, each element falls into one of $L$ 3-digit categories, where category $l$ is associated with $p_l \geq 1$ codes. The codes in category $l$ are denoted by $\boldsymbol{C}_{il}$.  The vector of coefficients for $\boldsymbol{C}_i$ has a special structure. It is partitioned into $\boldsymbol{\beta}_C = \{\boldsymbol{\beta}_{C_1}, \boldsymbol{\beta}_{C_2}, ... \boldsymbol{\beta}_{C_L}\}$ where $\boldsymbol{\beta}_{C_l}$ is the vector of coefficients for the grouped variables $\boldsymbol{C}_{il}$. To emphasize the grouped structure in the regression, we rewrite equation \ref{eqn:ps_model} as:
\begin{align}
\pi_i &= \mbox{logit}^{-1} \big\{\beta_0 + \boldsymbol{\beta}_B \boldsymbol{B}_i + \sum_{l=1}^L \sum_{k=1}^{p_l} \beta_{C_{l_k}} C_{il_k} \big\}. 
\end{align}

Capitalizing on the hierarchical structure, we define a hierarchical prior distribution as

\begin{align}
\boldsymbol{\beta}_{C_l} &\sim \mbox{Laplace}(\mu_l, .5) ~~\mbox{where}~~\mu_l \sim t_5(0, 2.5).
\end{align}

\noindent
The $t_5(0,2.5)$ priors on $\mu_l$ are again weakly informative. The 4-digit ICD-9 code coefficients are shrunk to the 3-digit (group) mean $\mu_l$ according to a Laplace distribution with a relatively tight scale. For example, \textit{a priori} each 4-digit code has a probability of 0.86 of being within 1 of the group mean coefficient $\mu_l$. The choice of the Laplace distribution over the Normal distribution forces noisier 4-digit codes to be shrunk more aggressively to the 3-digit group mean, while fatter tails permits less shrinking of signals \cite{gelman2014}. If there is only one 4-digit code in a 3-digit category, then we just assume $\beta_{C_l} = \mu_l$ and specify the weakly informative prior $\beta_{C_l} \sim t_5(0,2.5)$.

\begin{figure}[h]
	\begin{center}
		\includegraphics[trim = 0mm 0mm 0mm 0mm, clip, height=3.8in]{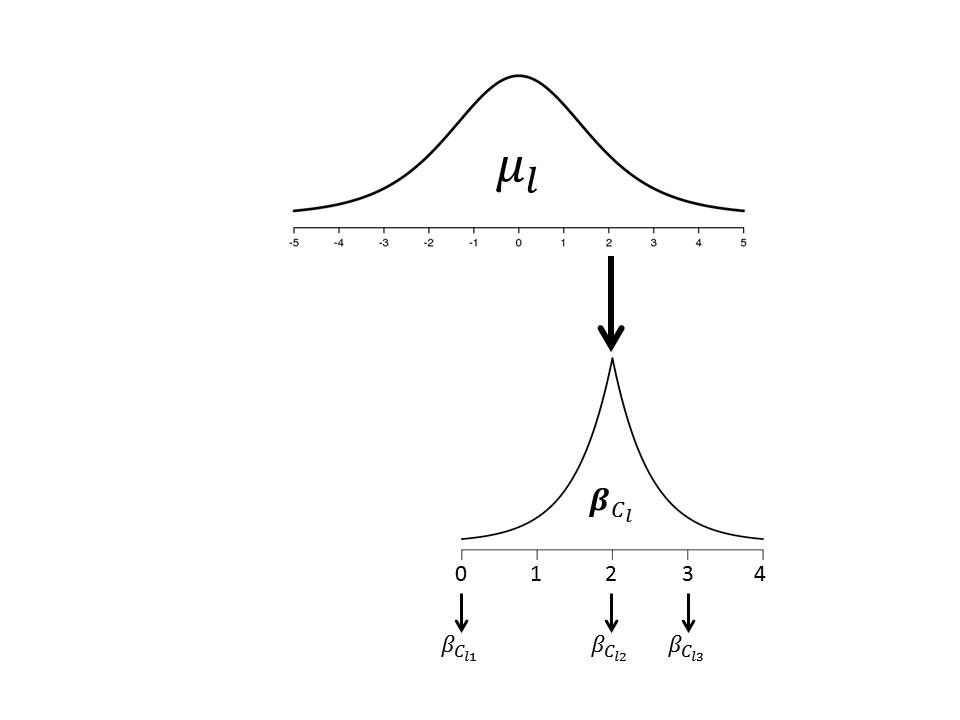}
		\caption{\textbf{Hierarchical prior on ICD-9 regression coefficients.} $t_5(0,2.5)$ prior distribution for $\mu_l$ (on top) is weakly informative for the group-level mean. $\boldsymbol{\beta}_{C_l}$, the coefficients for granular codes, are shrunk towards $\mu_l$ by a Laplace distribution.}
		\label{fig:model_schematic}
	\end{center}
\end{figure}

To fix ideas, imagine the group of ICD-9 codes corresponds to the category "hypertensive heart disease" and that it has an overall effect of $\mu_l = 2$ (Figure \ref{fig:model_schematic}). The 4-digit codes for this category include benign, unspecified, and malignant hypertensive heart disease. We might expect the benign condition to in fact have a small effect on stent selection such that $\beta_{C_{l_1}} = 0$; the malignant form of the condition may have a larger  effect than the group mean, so $\beta_{C_{l_3}} = 3$; and the unspecified condition is more ambiguous and receives an estimate equal to the group mean, $\beta_{C_{l_2}} = 2$. The fundamental advantage is that some information is shared among similar conditions, while different forms and severities are able to receive larger or smaller effect estimates as the data warrant. 

\subsection{Regularization via Comorbidity Scoring} 

Comorbidity indices are the results of a dimension reduction strategy that have the related advantages of low variance and parsimonious interpretation \cite{gilbert2017}. In typical usage, comorbidity indices compress hundreds or thousands of binary-valued ICD-9 codes into a smaller set of indicator variables or into a scalar summary, either of which can then be used for prediction or risk adjustment. Either way, the resulting claims data set is much smaller, with many of the original diagnosis data discarded or aggregated. The Elixhauser comorbidity index, for example, reduces all ICD-9 codes into 30 indicators, which can then be combined further into a scalar score using fixed severity weights \cite{elixhauser1998}\cite{walraven2009}. The severity weights are obtained from a regression model and are available in programs distributed by the developer of the index.  In our setting of causal inference, the association of the Elixhauser comorbidities are nuisance parameters -- their use is in meeting the ignorability of the treatment assignment assumption.

\section{Application}
\label{sec:massdac_analysis}

We implement a variety of methods to model propensity scores for DES vs BMS using the Bayesian propensity score approach to compare inferences using a variety of regularization strategies. We utilize ICD-9 diagnosis codes that are present-on-admission for a cohort of adults undergoing coronary stenting in hospitals.

\subsection{Claims Data and the Mass-DAC Registry}

We used claims data from the Massachusetts Center for Health Information and Analysis (CHIA). The CHIA data included 15 free-response fields with Present On Admission (POA) ICD-9 codes. Because POA codes describe patients at hospital arrival, they are not impacted by treatment decisions. To set a lower bound on sparsity we considered all 4-digit POA ICD-9 for which 10 or more patients were coded as having the condition.  

The claims data were supplemented with baseline demographic and clinical variables from the Mass-DAC registry, a state mandated database harvesting clinical information for all percutaneous coronary interventions (PCIs) performed in adults (age $\geq$ 18 years) in all non-Federal Massachusetts' hospitals annually. We identified 131 potential confounders in the Mass-DAC registry, including variables capturing demographic characteristics (e.g. age, race), pre-existing conditions (e.g. heart disease, diabetes), or procedure information (e.g. treated vessel, heart attack severity, hospital). The registry data are prospectively collected by hospital personnel who use the National Cardiovascular Data Registy instrument from the American College of Cardiology. A very small number of cells were missing data, and we imputed a single dataset for analysis.

We compared a number of outcomes between subjects implanted with a drug elusting stent (DES) and subjects implanted with bare metal stents (BMS) using Bayesian propensity score methods: 2-day mortality, 30-day mortality, 1-year mortality, and target vessel revascularization (TVR). Two-day and 30-day mortality are designed to serve as ``falsifiability endpoints." Because any difference in the stent performance would not manifest after such a short period, if we observe a difference on one of these outcomes it is almost certain it is due to residual confounding. We also generally expect no difference, or a very small difference, on 1-year mortality based on previous findings from RCTs and instrumental variable analyses \cite{bonaa2016}\cite{venkitachalam2011}. We do expect a significant benefit to DES on TVR rates. 

\subsection{Applied Methods}

In all methods, the Mass-DAC variables were included and their coefficients, $\boldsymbol{\beta}_B $, were specified with weakly informative $t_5(0,2.5)$ priors. Therefore only the modeling of claims data varied.

As baseline analyses, we fit two different Bayesian propensity score models to the log-odds of DES based on the Elixhauser index. We first grouped variables using the mapping of Quan et al from ICD-9 codes to the 30 original Elixhauser categories \cite{elixhauser1998}\cite{quan2005}. This mapping was accomplished in R using the package \texttt{medicalrisk} \cite{medicalrisk}. We then implemented a model with indicator variables for each category and used a weakly informative $t_5(0,2.5)$ prior. We refer to this model as ``Elixhauser indicator.'' Second, we replaced the 30 binary indicators with a comorbidity score created by summing the product of each indicator with its fixed severity weight.\cite{walraven2009} We refer to this model as the ``Elixhauser continuous.''

We used the full claims set (defined as all 4-digit POA ICD-9 for which 10 or more patients were coded as having the condition) with weakly informative, sparsity-inducing, and hierarchical priors to estimate the propensity score. As standard approaches a logistic regression using 3-digit codes and $t_5(0,2.5)$ priors and a logistic regression using 4-digit ICD-9 codes and $t_5(0,2.5)$ priors were fit \cite{normand1995}. Using 3-digit rather than 4-digit codes is a typical dimension reduction technique. We have found $t_5(0,2.5)$ priors to behave similarly to non-informative priors (analagous to maximum likelihood estimation) in this setting, but have slightly more posterior stability and easier sampling \cite{gelman2014}. As an alternative, we used horseshoe priors for strong regularization on the full 4-digit ICD-9 code set. Finally, we place a hierarchical prior on the coefficients so that coefficients associated with 4-digit codes are shrunk together towards the coefficients of the 3-digit codes.

The propensity score models were fit using Stan for Hamiltonian Monte Carlo sampling from each posterior \cite{stanjstatsoft}. The convergence of the samplers was assessed by examining trace plots and by the Gelman-Rubin ``r-hat" statistic, where a value less than 1.1 indicated sufficient convergence \cite{gelman2014}. To assess the out-of-sample fit of different models, we calculated the leave-one-out information criterion (LOO-IC). LOO-IC is an approximation to leave-one-out cross validation that can be computed from a point-wise log-likelihood matrix using Pareto-smoothed importance sampling, a procedure implemented in R through the \texttt{loo} package \cite{vehtari2017}\cite{loo}. A lower LOO-IC indicates a model with better predictive accuracy. 

To assess adjustment for measured confounders, we calculated how well the propensity scores balanced the observed covariates by calculating standardized differences, defined as the difference in means between the weighted BMS and DES populations over their pooled standard deviation. Because we have a different weighted population for each propensity score draw, this quantity is calculated once for each draw, giving a standardized difference distribution for each variable \cite{austin2015}. We assess the overall balance by determining the percentage of covariates with posterior mean standardized difference falling within (-10,10) \cite{austin2015}. Ideally, the 95\% intervals for standardized differences will also be contained within (-10,10) for most variables.
 
We assessed how well each method met our falsifiability endpoints and their causal effect estimates for 2-day, 30-day, and 1-year TVR. We were particularly interested in the coverage and width of the 95\% posterior intervals and whether they included zero.

\subsection{Results}

After eliminating diagnoses with fewer than 10 observed occurrences, we were left with a total of 334 4-digit ICD-9 confounders. There were fewer than 10 patients with individual codes in the ``psychoses'' Elixhauser category, so the Elixhauser indicator sets included 29 diagnoses rather than 30.

Grouping by the first 3-digits of the ICD-9 codes resulted in 69 groups containing at least 2 4-digit codes and 117 single 4-digit codes with no higher level group. The condition with the largest number of 4-digit codes in our sample was acute myocardial infarctions, i.e. heart attacks, and contained 10 4-digit codes. Thirty-seven of the 69 groups had only 2 4-digit codes. The complete list of all outcomes and confounders along with their prevalences (or means and standard deviations) appears in our appendix below.

%Thus the unadjusted differences (95\% CrI) between BMS and DES were -1.0\% (-1.4, -0.6) for 2-day mortalit y, -3.4\% (-4.1, -2.7) for 30-day mortality, -6.9\% (-8.0, -5.7) for 1-year mortality, and -2.4\% (-3.7, -1.3) for TVR.

%Figure \ref{fig:std_diff} shows the posterior mean and 95\% credible intervals for the standardized difference of each variable for each propensity score method. 

Table \ref{tab:prop_score} displays characteristics of the propensity score models including number of claims variables used, LOO-IC, and summaries of covariate balance. In terms of fit, Horseshoe regularization achieved the lowest LOO-IC score, indicating it had the best predictive accuracy. The Elixhauser indicator also had good LOO-IC, achieved through a considerable reduction in the dimension of the regression, though further reduction into a score degraded the quality of the fit considerably as indicated by a higher LOO-IC for Elixhauser continuous. The hierarchical prior provided lower LOO-IC than either the 4-digit or 3-digit $t_5(0,2.5)$ prior approaches. 

 In terms of balance, the Elixhauser methods and 3-digit codes with weakly informative priors failed to bring posterior mean standardized difference for all variables to within (-10,10). Regressions using the full claims set with weakly informative or hierarchical priors had many confounders for which the upper (or lower) 95\% posterior interval limit is beyond 10\% (or smaller than -10\%) after weighting. In contrast, the confounder balance when using a horseshoe prior results in only 6 variables with standardized difference posterior intervals outside the 10\% window.

\begin{table}
		 \resizebox{\textwidth}{!}{
		\begin{tabular}{l|cccc}
			\textbf{Approach} & \textbf{\# Claims Variables} & \textbf{LOO-IC (SE)} & \textbf{Means $\notin$ (-10,10)} & \textbf{CIs $\notin$ (-10, 10)} \\
			\hline
			\MyIndent Unadjusted$^*$ & 0 & 11327 (54) & 196 & NA \\
			
			\textit{Elixhauser Methods} & & & & \\ 
			
			\MyIndent $t_5(0,2.5)$ Prior Indicator Variables & 29 & 9537 (94)  & 3 & 11 \\
			\MyIndent $t_5(0,2.5)$ Prior Continuous Score & 1 & 9640 (92) & 4 & 12 \\
			
			\textit{Full Claims Set} & & & & \\
			
			\MyIndent $t_5(0,2.5)$ Prior 3-Digits & 186 & 9662 (100) & 1 & 59 \\
			\MyIndent $t_5(0,2.5)$ Prior 4-Digits & 334 & 9741 (107) & 0 & 150 \\
			\MyIndent Horseshoe Prior 4-Digits & 334 & 9468 (88) & 0 & 6  \\
			\MyIndent Hierarchical Prior 3 and 4-Digits & 334 & 9639 (104)  & 0 & 74\\
		\end{tabular}
}
		\caption{\textbf{Characteristics of propensity score models.} Number of claims derived variables used in model, leave-one-out information criterion, and summaries of covariate balance for unadjusted analysis and each propensity score analysis. Column ``Means $\notin$ (-10,10)" tabulates the number of variables with posterior mean standardized differences outside the interval -10 to 10, which is typically considered good balance. Column ``CIs $\notin$ (-10, 10)" tabulates number of variables where some part of the 95\% posterior interval lies outside (-10,10). LOO-IC = leave-one-out information criterion; SE = standard error. $^*$: This is a model with an intercept only for the purpose of calculating LOO-IC and standardized differences reflect covariates in unweighted population.}
		\label{tab:prop_score}

\end{table}

\begin{figure}[h]
	\begin{center}
		\includegraphics[trim = 0mm 0mm 0mm 0mm, clip, width = .49\textwidth]{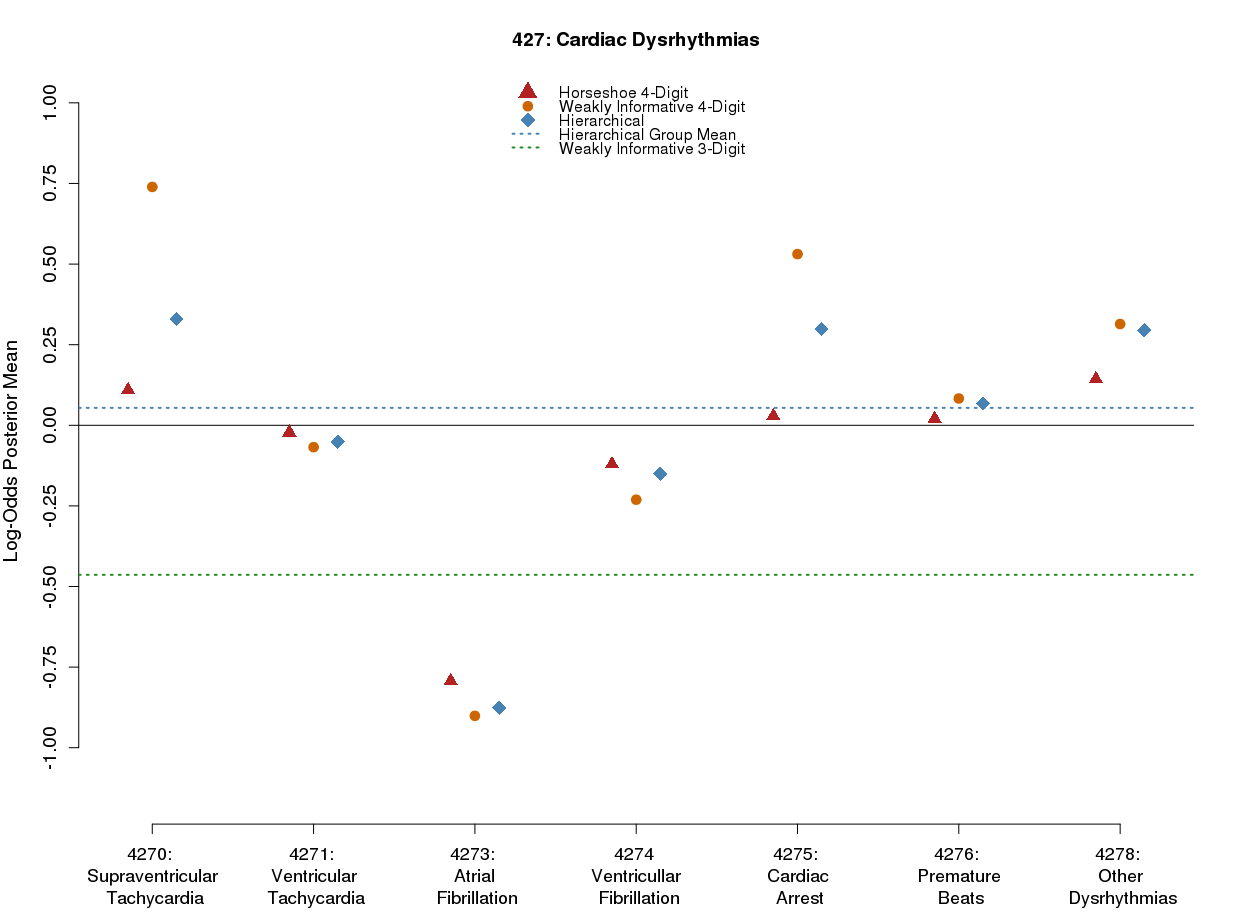}
		\includegraphics[trim = 0mm 0mm 0mm 0mm, clip, width = .49\textwidth]{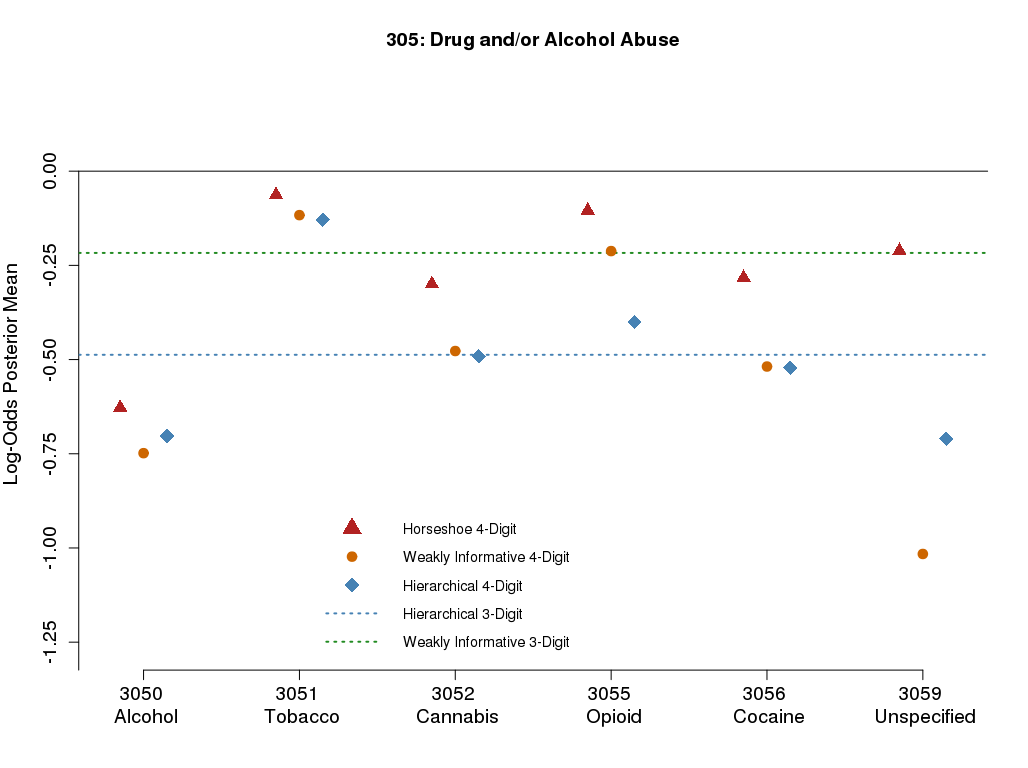}
		\caption{\textbf{Selected posterior mean coefficients.} Values on log-odds scale for 4-digit ICD-9 coefficients  with weakly informative priors (orange circle), horseshoe (red triangle), and hierarchical (blue diamond) specifications. Group mean defined by 3-digit code for hierarchical (blue dashed line) and 3-digit weakly informative coefficient estimate (green dashed line) are also shown. Left panel shows coefficients for code group 427, cardiac dysrhythmias; right panel shows coefficients for code group 305, drug or alcohol abuse.}
		\label{fig:coefficients}
	\end{center}
\end{figure}

Figure \ref{fig:coefficients} illustrates the behavior of the different approaches to regularization when using the full claims set for two groups (e.g., two 3-digit diagnosis codes): cardiac dysrhythmias and drug/alcohol abuse. For cardiac dysrhythmias, the group mean for the hierarchical prior coefficients is very close to zero because the associations of the individual codes have opposite signs on the logit of the probability of receiving a DES. The hierarchical prior coefficients are slightly pooled towards their group mean, while coefficients under the horseshoe are pulled towards zero, in some cases quite sharply. The weakly informative coefficients are universally larger in magnitude than the others. A similar pattern emerges for drug/alcohol abuse coefficients. The group overall is associated with a lower probability of receiving a DES, with some granular codes exhibiting much smaller or larger effects. For both coefficient groups, a weakly informative prior using a single indicator for the 3-digit code category fails to capture the nuances apparent at lower levels.

Table \ref{tab:massdac_results} presents posterior summaries of causal effects in the Mass-DAC data under each method. The unadjusted difference in means between the two stent groups exhibits considerable confounding of both 30-day and 1-year mortality with large risk reductions for DES of 3.4\% and 6.9\% respectively. The Elixhauser approaches were the only models unable to adjust away the difference in 2-day mortality, though margins were slim. For 30-day mortality, only the full claims sets with $t_5(0,2.5)$ or hierarchical priors estimated no significant effect. In contrast, aggressively regularizing using horseshoe priors yielded a 30-day mortality benefit to DES that is indicative of residual confounding.  On one-year mortality, the model utilizing the hierarchical prior distribution for the ICD-9 coefficients achieved a considerable reduction from the large unadjusted difference, and improved over the comparably large effect estimates generated from the Elixhauser and horseshoe prior approaches. However no method included zero in its 95\% credible interval, conflicting with findings from randomized trials. As expected, all methods produced a larger benefit to TVR than indicated by the unadjusted difference.

The Elixhauser indicator, Elixhauser continuous, and horseshoe approaches had the lowest uncertainty with posterior interval widths of 1.9, 1.8, and 2.0 respectively. The weakly informative 3-digit approach had an average width of 2.3, while the weakly informative 4-digit approach had an average of width of 3.0. The hierarchical prior approach had an average credible interval width of 2.5, about 20\% tighter than the 4-digit approach. The weakly informative and hierarchical specifications thus incur more variance than the smaller or more regularized Elixhauser and horseshoe approaches, which is ultimately reflected in the estimated distributions of causal parameters.

\begin{table}[h]
	 \resizebox{\textwidth}{!}{
	\begin{tabular}{l|cccc}
		& \multicolumn{3}{c}{\textbf{Mortality}} & \\
		\cline{2-4}
		\textbf{Approach} &  \textbf{2-Day} & \textbf{30-Day} & \textbf{1-Year} & \textbf{1-Year TVR}\\
		\hline
		\MyIndent Unadjusted &  -1.0 (-1.4, -0.6) & -3.4 (-4.1, -2.7) & -6.9 (-8.0, -5.7) & -2.4 (-3.7, -1.3) \\
		
		\textit{Elixhauser Methods} & & & &\\ 
	
		\MyIndent $t_5(0,2.5)$ Prior Indicator Variables & -0.4 (-0.7, -0.1) & -1.2 (-2.0, -0.5) & -3.1 (-4.3, -1.8) & -4.3 (-5.7, -2.9) \\
		\MyIndent $t_5(0,2.5)$ Prior Continuous Score & -0.4 (-0.7, -0.1) & -1.3 (-2.0, -0.6) & -3.3 (-4.5, -2.1) & -4.5 (-5.9, -3.0)\\
		
		\textit{Full Claims Set} & & & & \\
		
	    \MyIndent $t_5(0,2.5)$ Prior 3-Digits & -0.3 (-0.7, 0.1) & -1.0 (-2.0, 0.2) & -2.7 (-4.1, -1.1) & -4.4 (-6.1,-2.8) \\
		\MyIndent $t_5(0,2.5)$ Prior 4-Digits &  -0.3 (-0.7, 0.0) & -0.7 (-2.0, 1.2) & -2.5 (-4.5, -0.3) & -3.6 (-5.5, -1.8)\\
		\MyIndent Horseshoe Prior 4-Digits & -0.3 (-0.7, 0.0) & -1.1 (-1.9, -0.3) & -3.0 (-4.3, -1.7) & -4.1 (-5.6, -2.6)\\
		\MyIndent Hierarchical Prior 3 and 4-Digits & -0.3 (-0.7, 0.1) & -0.9 (-2.0, 0.5) & -2.6 (-4.2, -0.8) & -3.8 (-5.5, -2.0) \\
		%\MyIndent Bayes Lasso Hierarchical & & 9390 (88) & -0.3 (-0.7, 0.0) & -1.3 (-2.0, -0.5) & -3.2 (-4.4, -2.0) & -3.9 (-5.3, -2.6) \\
	\end{tabular}
}
	\caption{\textbf{Casual effect estimates for selected outcomes.} Risk differences for various methods applied to the coronary stent study. Risk differences (95\% Credible Intervals) estimated from the Mass-DAC data show benefit of DES compared to BMS in terms of reduced percent chance of given outcome ($\Delta = p_{\mbox{\tiny DES}} - p_{\mbox{\tiny BMS}}$). TVR = target vessel revascularization.}
	\label{tab:massdac_results}
\end{table}

\section{Discussion}

In this paper, we proposed the use of regularizing prior distributions rather than comorbidity indices for data reduction strategies in high-dimensional causal inference problems in health care. We found that Bayesian priors improved inferences over comorbidity indices both in terms of out-of-sample fit  or the ability to meet falsifiability endpoints. These prior distributions are able to work with a large number of claims variables while controlling variance. 

To compare multiple methods with and without hierarchical modeling of ICD-9 codes, we estimated causal effects of DES compared to BMS on a number of outcomes. Specifically, we were looking for our methods to adjust away any difference in 30-day mortality and, ideally, 1-year mortality as well. RCTs have consistently shown no survival benefit to DES, but this has proved to be very difficult to replicate in observational analyses \cite{bonaa2016}\cite{venkitachalam2011}\cite{stettler2007}\cite{mauri_circ2008}. By using claims data and either a hierarchical ICD-9 or flat weakly informative propensity score model, we succeeded in removing the specious benefit shown to 2-day and 30-day mortality, which were very likely to be caused by residual confounding. In contrast, models using Elixhauser comorbidity indicators, a continuous Elixhauser score, or sparsity-inducing horseshoe priors on the full claims set did not manage to adjust away confounding on 30-day mortality. The fact that these dimension reducing approaches generated the best out-of-sample fit in terms of LOO-IC indicates the importance of using additional checks like falsifiability endpoints in causal inference problems, rather than relying solely on scores of predictive ability. 

Unfortunately no method succeeded in removing all confounding of 1-year mortality, but the hierarchical and weakly informative prior distributions had comparably more success, estimating small effect sizes and 95\% credible intervals close to 0. Residual confounding has plagued past observational studies of DES and BMS and it appears that, while claims data are able to account for important confounders not found in registries, we are still lacking the necessary variables to completely remove confounding \cite{mauri_circ2008}. Likely unmeasured confounders include socio-economic variables, which past studies have shown affect adherence to dual-antiplatelet therapy and ultimately a doctor's decision to use a DES versus a BMS in usual care settings \cite{spertus2006}.

While generating similar posterior means, the hierarchical prior approach had considerably smaller variance than the unregularized approach, with about 20\% tighter credible intervals on average. All methods showed, as expected, a considerable benefit to DES on the TVR outcome.    

We restricted our attention to Bayesian propensity scores throughout, only changing the variable selection or regularization strategy for comparability. However, hierarchical ICD-9 code modeling is not limited to Bayesian propensity scores and could easily be adapted to Bayesian outcome modeling or frequentist techniques. Whatever the specifics, our work demonstrates that investigators performing causal inference in non-randomized settings could benefit by ignoring comorbidity indices and using more claims variables in their analyses. Comorbidity indices will continue to have utility for risk stratification and parsimonious assessments in clinical practice, but as we have shown, there are better alternatives for confounder adjustment.

\section{Acknowledgments}
Mr Spertus and Dr Normand were supported by a grant from the US Food and Drug Administration (U01FD004493). Mr Spertus, Dr Normand, and Dr Adhikari were supported by a grant from the National Institute of General Medical Sciences (1R01GM111339).

We are indebted to the Massachusetts Department of Public Health (MDPH) and the Massachusetts Center for Health Information and Analysis Case Mix Databases (CHIA) for the use of their data.

\bibliography{CERinBigdataWorksCited}

\appendix
\section{Covariate Tables}

\subsection{Registry Covariates}

\begin{longtable}{lrrr}
	\caption{Prevalence or mean (SD) of outcomes in the Mass-DAC registry. ``Cases per Site" gives min, mean (SD), and max of numbers treated at the 25 hospitals, overall and in each treatment group. The 31 coronary artery segment indicators are not shown. HMO = health maintenance organization; CAD = coronary artery disease; CVD = cardiovascular disease; PAD = peripheral artery disease; CABG = coronary artery bypass graft; STEMI = ST-elevated myocardial infarction; LVSD = left ventricular systolic dysfunction; NYHA = New York heart association; LAD = left anterior descending (artery); RCA = right coronary artery.}\\
	\hline
	& \textbf{Overall}  & \textbf{BMS}  & \textbf{DES} \\ 
	\hline
	Number Treated ($n$) & 8718 & 3081 & 5637 \\
	\hline
	\textit{\textbf{Registry Outcomes}} & & & \\
	\% 2 Day Mortality & 0.50 & 1.14 & 0.16 \\
	\% 30 Day Mortality & 2.01 & 4.19 & 0.82 \\  
	\% 1 Year TV Revascularization & 7.38 & 8.96 & 6.51 \\  
	\% 1 Year Mortality & 5.75 & 10.19 & 3.32 \\ 
	
	\hline
	\textit{\textbf{Registry Confounders}} \\
	\% Male & 69.39 & 68.16 & 70.05 \\ 
	Mean (SD) Age in Years  & 64.66 (12.5) & 66.37 (11.7) & 63.72 (13.5) \\ 
	Mean (SD) Height in cm  & 170.67 (10.5) & 170.41 (10.6) & 170.81 (10.4) \\ 
	Mean (SD) Weight in kg  & 86.48 (20.3) & 84.89 (19.9) & 87.34 (20.8) \\
	
	\rowgroup{Race}\\
	\MyIndent \% Caucasian & 91.26 & 90.72 & 91.56 \\ 
	\MyIndent \% Black & 3.79 & 4.87 & 3.19 \\ 
	\MyIndent \% Asian & 2.31 & 2.43 & 2.24 \\ 
	\MyIndent \% Native American & $<0.1$ & $<0.3$& $<0.2$ \\ 
	\MyIndent \% Native Pacific & $<0.1$& $<0.3$& $<0.2$\\ 
	
	\% Hispanic or Latino  & 4.55 & 4.8 & 4.42 \\ 
	
	\rowgroup{Payor}\\
	\MyIndent \% Government & 51.57 & 59.17 & 47.42\\
	\MyIndent \% None & 2.56 & 4.12 & 1.7 \\ 
	\MyIndent \% Non-US & $<0.1$ & $<0.3$ & $<0.2$ \\ 
	\MyIndent \% Private Commercial or HMO & 45.82 & 36.68 & 50.82 \\ 
	
	\% Smoker & 74.68 & 70.85 & 76.78 \\ 
	\% Hypertension & 78.3 & 76.6 & 79.23 \\ 
	\% Dyslipidemia & 81.82 & 77.96 & 83.93 \\ 
	\% Diabetes & 32.16 & 30.38 & 33.14 \\ 
	\% Family History CAD & 28.96 & 24.08 & 31.63 \\ 
	\% Chronic Lung Disease & 14.21 & 16.42 & 13 \\ 
	\% Current Dialysis & 1.82 & 2.21 & 1.61 \\ 
	
	\% Prior CVD & 10.23 & 11.33 & 9.63 \\ 
	\% Prior PAD & 11.96 & 12.76 & 11.53 \\ 
	\% Prior Myocardial Infarction & 26.67 & 25.84 & 27.12 \\ 
	\% Prior Heart Failure & 10.61 & 12.46 & 9.6 \\ 
	\% Prior Valve Surgery & 1.56 & 2.34 & 1.14 \\ 
	\% Prior PCI & 27.56 & 19.77 & 31.83 \\ 
	\% Prior CABG & 12.3 & 11.46 & 12.76 \\ 
	\% Prior Cardiogenic Shock & 1.9 & 3.89 & 0.82 \\ 
	\% Prior Cardiac Arrest & 2.29 & 4.15 & 1.28 \\ 
	
	\rowgroup{CAD Presentation}\\
	\MyIndent \% No Angina & 5.32 & 5.78 & 5.07 \\ 
	\MyIndent \% Symptom Unlikely to be Ischemic & 1.00 & 1.17 & 0.90 \\ 
	\MyIndent \% Stable Angina & 11.88 & 5.81 & 15.2 \\ 
	\MyIndent \% Unstable Angina & 30.19 & 23.4 & 33.9 \\ 
	\MyIndent \% Non-STEMI & 27.23 & 28.17 & 26.72 \\ 
	\MyIndent \% STEMI & 24.37 & 35.67 & 18.2 \\ 
	
	\% Thrombolytic Therapy & 0.8 & 1.23 & 0.57 \\ 
	\% Cardiomyopathy or LVSD & 9.36 & 11.1 & 8.41 \\ 
	
	\rowgroup{Anginal Canadian Classification}\\
	\MyIndent \% 0 & 9.64 & 11.65 & 8.53\\
	\MyIndent \% I & 2.4 & 1.59 & 2.84 \\ 
	\MyIndent \% II & 12.2 & 7.66 & 14.69 \\ 
	\MyIndent \% III & 28.83 & 24.15 & 31.38 \\ 
	\MyIndent \% IV & 46.94 & 54.95 & 42.56 \\ 
	
	\rowgroup{Anti-Anginal Medications}\\
	\MyIndent \% Beta Blockers & 57.79 & 54.56 & 59.55 \\ 
	\MyIndent \% Calcium Channel Blockers & 11.44 & 11.39 & 11.46 \\ 
	\MyIndent \% Long Acting Nitrates & 11.49 & 9.41 & 12.63 \\ 
	\MyIndent \% Ranolazine & 0.62 & 0.45 & 0.71 \\ 
	\MyIndent \% Other Agent & 1.4 & 1.01 & 1.61 \\ 
	
	\rowgroup{NYHA Class}\\
	\MyIndent \% 0 & 87.57 & 83.51 & 89.78\\
	\MyIndent \% I & 0.91 & 0.78 & 0.98 \\ 
	\MyIndent \% II & 3.06 & 3.67 & 2.73 \\ 
	\MyIndent \% III & 4.16 & 5.39 & 3.49 \\ 
	\MyIndent \% IV & 4.3 & 6.65 & 3.02 \\

	\% Compassionate Use & 0.91 & 2.01 & 0.3 \\ 
	\% Cardiogenic Shock & 1.85 & 3.8 & 0.78 \\ 
	\% Mechanical Ventricular Support & 0.58 & 1.14 & 0.28 \\ 
	\% Ejection Fraction $<30\%$ & 2.88 & 3.7 & 2.43 \\ 
	
	\rowgroup{Coronary Anatomy}\\
	\MyIndent \% Left Dominant & 8.30 & 8.08 & 8.42 \\
	\MyIndent \% Right Dominant & 86.19 & 85.88 & 86.36 \\ 
	\MyIndent \% Left Dominant & 5.51 & 6.04 & 5.22 \\ 
	
	\% Left Main Disease & 5.93 & 6.52 & 5.61 \\ 
	Mean Left Main Stenosis (SD) & 7.98 (19.8) & 8.74(19.3) & 7.57(20.7) \\ 
	Mean Proximal LAD Stenosis (SD)& 36 (39.7) & 35.1(39.7) & 36.5(39.5) \\ 
	Mean Mid-Distal LAD Stenosis (SD)& 48.2 (39.7) & 47.64(39.7) & 48.51(39.6) \\ 
	Mean Circumflex Stenosis (SD)& 47.95 (40.8) & 47.94(40.7) & 47.96(40.8) \\ 
	Mean RCA Stenosis (SD)& 60.66 (39.9) & 65.25(40) & 58.15(39.2) \\ 
	
	\rowgroup{Status}\\
	\MyIndent \% Urgent & 52.49 & 48.78 & 54.51 \\ 
	\MyIndent \% Emergent & 26.69 & 38.30 & 20.35 \\ 
	\MyIndent \% Other & 20.82 & 12.92 & 25.14\\
	
	\rowgroup{PCI Indication}\\
	\MyIndent \% Immediate PCI for STEMI & 21.80 & 31.39 & 16.55\\
	\MyIndent \% STEMI (Unstable, $>12$ hours) & 1.67 & 2.66 & 1.14 \\ 
	\MyIndent \% STEMI (Stable, $>12$ hours) & 0.57 & 0.94 & 0.37 \\ 
	\MyIndent \% STEMI (Stable, thrombolytics) & 0.44 & 0.58 & 0.35 \\ 
	\MyIndent \% STEMI (Rescue, failed thrombolytics) & 0.46 & 0.75 & 0.3 \\ 
	\MyIndent \% High risk Non-STEMI & 48.99 & 45.86 & 50.7 \\ 
	\MyIndent \% Staged & 0.86 & 0.45 & 1.08 \\ 
	\MyIndent \% Other & 25.21 & 17.36 & 29.5 \\ 
	
	Mean (SD) Thrombectomies & 0.12 (0.3) & 0.16(0.3) & 0.09(0.4) \\
	Mean (SD) Total Stents Used & 1.46 (0.8) & 1.41 (0.8) & 1.5 (0.8) \\  
	Mean (SD) Lesions & 1.3 (0.6) & 1.24 (0.6) & 1.33 (0.5) \\
	Mean (SD) Lesion Length & 18.08 (10) & 17 (10.4) & 18.68 (9.3) \\ 
	\% In-stent Restenosis & 6.96 & 2.34 & 9.49 \\ 
	\% Chronic Total Occlusion & 1.73 & 1.3 & 1.97 \\ 
	
	\rowgroup{Cases per Site (25 Total)}\\
	\MyIndent  Min \# Treated & 47 & 12 & 14\\
	\MyIndent  Mean \# Treated (SD) & 349 (269) & 123 (99) & 225 (187)\\
	\MyIndent  Max \# Treated & 896 & 352 & 605 \\

	\hline
	
	\label{tab:mass_dac_covariates} 
\end{longtable}

\newpage

% latex table generated in R 3.4.0 by xtable 1.8-2 package
% Fri Aug  4 14:02:17 2017
\begin{longtable}{llr}
	\caption{Prevalence of present on admission ICD-9 diagnosis codes used in analysis. As with the analysis using a hierarchical prior, codes are grouped together if there were two or more 4-digit codes sharing a 3-digit grouping. w/o = without.}\\
	\centering
	Diagnosis Code & Description (Present on Admission) & Prevalence \\ 
	\hline
	041 & Bacterial infection of unspecified site & 0.5 \\ 
	\MyIndent 0411 & Staphylococcus infection, unspecified & 0.2 \\ 
	\MyIndent0414 & Shiga toxin-producing E. coli (STEC) & 0.3 \\ 
	070 & Viral hepatitis & 1.1 \\ 
	\MyIndent 0703 & Viral hepatitis B w/o mention of hepatic coma & 0.2 \\ 
	\MyIndent 0705 & Acute hepatitis C w/o mention of hepatic coma & 0.4 \\ 
	\MyIndent 0707 & Unspecified viral hepatitis C w/o hepatic coma & 0.5 \\ 
	242 & Thyrotoxicosis with or w/o goiter & 0.3 \\ 
	\MyIndent 2420 & Toxic diffuse goiter w/o mention of thyrotoxic crisis or storm & 0.1 \\ 
	\MyIndent 2429 & Thyrotoxicosis w/o mention of goiter, thyrotoxic crisis, or storm & 0.2 \\ 
	244 & Acquired hypothyroidism & 7.4 \\ 
	\MyIndent 2440 & Postsurgical hypothyroidism & 0.2 \\ 
	\MyIndent 2448 & Other specified acquired hypothyroidism & 0.2 \\ 
	\MyIndent 2449 & Unspecified acquired hypothyroidism & 7 \\ 
	250 & Diabetes mellitus & 30.8 \\ 
	\MyIndent 2500 & Diabetes mellitus w/o mention of complication & 27 \\ 
	\MyIndent 2501 & Diabetes with ketoacidosis & 0.2 \\ 
	\MyIndent 2504 & Diabetes with renal manifestations & 1.5 \\ 
	\MyIndent 2505 & Diabetes with ophthalmic manifestations & 1.1 \\ 
	\MyIndent 2506 & Diabetes with neurological manifestations & 1.9 \\ 
	\MyIndent 2507 & Diabetes with peripheral circulatory disorders & 0.2 \\ 
	\MyIndent 2508 & Diabetes with other specified manifestations & 0.3 \\ 
	272 & Disorders of lipoid metabolism & 69.9 \\ 
	\MyIndent 2720 & Pure hypercholesterolemia & 13.7 \\ 
	\MyIndent 2721 & Pure hyperglyceridemia & 1 \\ 
	\MyIndent 2722 & Mixed hyperlipidemia & 0.1 \\ 
	\MyIndent 2724 & Other and unspecified hyperlipidemia & 59.4 \\ 
	\MyIndent 2729 & Unspecified disorder of lipoid metabolism & 0.1 \\ 
	274 & Gout & 3.5 \\ 
	\MyIndent 2740 & Gouty arthropathy, unspecified & 0.3 \\ 
	\MyIndent 2749 & Gout, unspecified & 3.2 \\ 
	275 & Disorders of mineral metabolism & 0.9 \\ 
	\MyIndent 2750 & Hereditary hemochromatosis & 0.1 \\ 
	\MyIndent 2752 & Disorders of magnesium metabolism & 0.3 \\ 
	\MyIndent 2753 & Disorders of phosphorus metabolism & 0.2 \\ 
	\MyIndent 2754 & Unspecified disorder of calcium metabolism & 0.3 \\ 
	276 & Disorders of fluid, electrolyte, and acid-base balance & 4.5 \\ 
	\MyIndent 2760 & Hyperosmolality and/or hypernatremia & 0.1 \\ 
	\MyIndent 2761 & Hyposmolality and/or hyponatremia & 1.2 \\ 
	\MyIndent 2762 & Acidosis & 0.7 \\ 
	\MyIndent 2764 & Mixed acid-base balance disorder & 0.1 \\ 
	\MyIndent 2765 & Volume depletion, unspecified & 0.7 \\ 
	\MyIndent 2766 & Transfusion associated circulatory overload & 0.1 \\ 
	\MyIndent 2767 & Hyperpotassemia & 0.7 \\ 
	\MyIndent 2768 & Hypopotassemia & 1.1 \\ 
	280 & Iron deficiency anemias & 1.4 \\ 
	\MyIndent 2800 & Iron deficiency anemia secondary to blood loss (chronic) & 0.2 \\ 
	\MyIndent 2809 & Iron deficiency anemia, unspecified & 1.2 \\ 
	281 & Other deficiency anemias & 0.3 \\ 
	\MyIndent 2810 & Pernicious anemia & 0.1 \\ 
	\MyIndent 2819 & Unspecified deficiency anemia & 0.1 \\ 
	285 & Other and unspecified anemias & 5.5 \\ 
	\MyIndent 2851 & Acute posthemorrhagic anemia & 0.1 \\ 
	\MyIndent 2852 & Anemia in chronic kidney disease & 1.5 \\ 
	\MyIndent 2858 & Other specified anemias & 0.1 \\ 
	\MyIndent 2859 & Anemia, unspecified & 3.8 \\ 
	287 & Purpura and other hemorrhagic conditions & 1.3 \\ 
	\MyIndent 2874 & Posttransfusion purpura & 0.1 \\ 
	\MyIndent 2875 & Thrombocytopenia, unspecified & 1.2 \\ 
	294 & Other organic psychotic conditions (chronic) & 1 \\ 
	\MyIndent 2941 & Dementia  w/o behavioral disturbance & 0.2 \\ 
	\MyIndent 2948 & Other persistent mental disorders & 0.8 \\ 
	296 & Affective psychoses & 1.4 \\ 
	\MyIndent 2962 & Major depressive affective disorder, single episode & 0.2 \\ 
	\MyIndent 2963 & Major depressive affective disorder, recurrent episode & 0.1 \\ 
	\MyIndent 2968 & Bipolar disorder, unspecified & 0.9 \\ 
	\MyIndent 2969 & Unspecified episodic mood disorder & 0.1 \\ 
	300 & Neurotic disorders & 6.8 \\ 
	\MyIndent 3000 & Anxiety state, unspecified & 4.6 \\ 
	\MyIndent 3004 & Dysthymic disorder & 2.2 \\ 
	305 & Nondependent abuse of drugs & 21.9 \\ 
	\MyIndent 3050 & Alcohol abuse, unspecified & 2.2 \\ 
	\MyIndent 3051 & Tobacco use disorder & 20.4 \\ 
	\MyIndent 3052 & Cannabis abuse, unspecified & 0.9 \\ 
	\MyIndent 3055 & Opioid abuse, unspecified & 0.2 \\ 
	\MyIndent 3056 & Cocaine abuse, unspecified & 0.7 \\ 
	\MyIndent 3059 & Other, mixed, or unspecified drug abuse, unspecified & 0.2 \\ 
	333 & Extrapyramidal disease and abnormal movement disorders & 0.6 \\ 
	\MyIndent 3331 & Essential and other specified forms of tremor & 0.2 \\ 
	\MyIndent 3339 & Unspecified extrapyramidal disease and abnormal movement & 0.4 \\ 
	338 & Pain, not elsewhere classified & 2 \\ 
	\MyIndent 3382 & Chronic pain due to trauma & 1.8 \\ 
	\MyIndent 3384 & Chronic pain syndrome & 0.2 \\ 
	348 & Other conditions of brain & 0.5 \\ 
	\MyIndent 3481 & Anoxic brain damage & 0.4 \\ 
	\MyIndent 3483 & Encephalopathy, unspecified & 0.1 \\ 
	355 & Mononeuritis of lower limb & 0.4 \\ 
	\MyIndent 3558 & Mononeuritis of lower limb, unspecified & 0.1 \\ 
	\MyIndent 3559 & Mononeuritis of unspecified site & 0.3 \\ 
	362 & Other retinal disorders & 1.4 \\ 
	\MyIndent 3620 & Background diabetic retinopathy & 1 \\ 
	\MyIndent 3625 & Macular degeneration (senile), unspecified & 0.4 \\ 
	396 & Diseases of mitral and aortic valves & 0.7 \\ 
	\MyIndent 3962 & Mitral valve insufficiency and aortic valve stenosis & 0.3 \\ 
	\MyIndent 3963 & Mitral valve insufficiency and aortic valve insufficiency & 0.3 \\ 
	401 & Essential hypertension & 62.4 \\ 
	\MyIndent 4010 & Malignant essential hypertension & 0.1 \\ 
	\MyIndent 4011 & Benign essential hypertension & 0.7 \\ 
	\MyIndent 4019 & Unspecified essential hypertension & 61.6 \\ 
	410 & Acute myocardial infarction & 54.3 \\ 
	\MyIndent 4100 & AMI of anterolateral wall & 2.7 \\ 
	\MyIndent 4101 & AMI of other anterior wall & 6.4 \\ 
	\MyIndent 4102 & AMI of inferolateral wall & 2.2 \\ 
	\MyIndent 4103 & AMI of inferoposterior wall & 2.4 \\ 
	\MyIndent 4104 & AMI of other inferior wall & 9.5 \\ 
	\MyIndent 4105 & AMI of other lateral wall & 0.9 \\ 
	\MyIndent 4106 & True posterior wall infarction & 0.2 \\ 
	\MyIndent 4107 & Subendocardial infarction & 28.7 \\ 
	\MyIndent 4108 & AMI of other specified sites & 0.3 \\ 
	\MyIndent 4109 & AMI of unspecified site & 1.1 \\ 
	411 & Other acute and subacute form of ischemic heart disease & 19.8 \\ 
	\MyIndent 4111 & Intermediate coronary syndrome & 19.3 \\ 
	\MyIndent 4118 & Acute coronary occlusion w/o myocardial infarction & 0.5 \\ 
	414 & Other forms of chronic ischemic heart disease & 86.7 \\ 
	\MyIndent 4140 & Coronary atherosclerosis of unspecified vessel & 85.6 \\ 
	\MyIndent 4141 & Aneurysm of heart (wall) & 0.6 \\ 
	\MyIndent 4142 & Chronic total occlusion of coronary artery & 9.5 \\ 
	\MyIndent 4148 & Other specified forms of chronic ischemic heart disease & 4.9 \\ 
	\MyIndent 4149 & Chronic ischemic heart disease, unspecified & 0.2 \\ 
	424 & Other diseases of endocardium & 6.3 \\ 
	\MyIndent 4240 & Mitral valve disorders & 3.5 \\ 
	\MyIndent 4241 & Aortic valve disorders & 2.8 \\ 
	425 & Cardiomyopathy & 2.6 \\ 
	\MyIndent 4254 & Other primary cardiomyopathies & 2.5 \\ 
	\MyIndent 4255 & Alcoholic cardiomyopathy & 0.1 \\ 
	426 & Conduction disorders & 5.2 \\ 
	\MyIndent 4260 & Atrioventricular block, complete & 0.8 \\ 
	\MyIndent 4261 & Atrioventricular block, unspecified & 1.3 \\ 
	\MyIndent 4262 & Left bundle branch hemiblock & 0.1 \\ 
	\MyIndent 4263 & Other left bundle branch block & 1.4 \\ 
	\MyIndent 4264 & Right bundle branch block & 1.5 \\ 
	\MyIndent 4265 & Bundle branch block, unspecified & 0.4 \\ 
	427 & Cardiac dysrhythmias & 17.3 \\ 
	\MyIndent 4270 & Paroxysmal supraventricular tachycardia & 0.2 \\ 
	\MyIndent 4271 & Paroxysmal ventricular tachycardia & 2.4 \\ 
	\MyIndent 4273 & Atrial fibrillation & 10 \\ 
	\MyIndent 4274 & Ventricular fibrillation & 1.6 \\ 
	\MyIndent 4275 & Cardiac arrest & 1.3 \\ 
	\MyIndent 4276 & Premature beats, unspecified & 0.7 \\ 
	\MyIndent 4278 & Sinoatrial node dysfunction & 4.3 \\ 
	428 & Heart failure & 15.5 \\ 
	\MyIndent 4280 & Congestive heart failure, unspecified & 14.5 \\ 
	\MyIndent 4282 & Systolic heart failure, unspecified & 5.6 \\ 
	\MyIndent 4283 & Diastolic heart failure, unspecified & 3 \\ 
	\MyIndent 4284 & Combined systolic and diastolic heart failure, unspecified & 1.6 \\ 
	429 & Ill-defined descriptions and complications of heart disease & 1.8 \\ 
	\MyIndent 4293 & Cardiomegaly & 0.7 \\ 
	\MyIndent 4298 & Other disorders of papillary muscle & 0.2 \\ 
	\MyIndent 4299 & Heart disease, unspecified & 0.9 \\ 
	433 & Occlusion and stenosis of precerebral arteries & 1.8 \\ 
	\MyIndent 4331 & Occlusion/stenosis of carotid artery & 1.8 \\ 
	\MyIndent 4333 & Occlusion/stenosis of multiple precerebral arteries & 0.5 \\ 
	440 & Atherosclerosis & 2.6 \\ 
	\MyIndent 4400 & Atherosclerosis of aorta & 0.3 \\ 
	\MyIndent 4401 & Atherosclerosis of renal artery & 0.4 \\ 
	\MyIndent 4402 & Atherosclerosis of native arteries of the extremities & 1.9 \\ 
	\MyIndent 4408 & Atherosclerosis of other specified arteries & 0.2 \\ 
	441 & Aortic aneurysm and dissection & 1.2 \\ 
	\MyIndent 4412 & Thoracic aneurysm w/o mention of rupture & 0.1 \\ 
	\MyIndent 4414 & Abdominal aneurysm w/o mention of rupture & 1.1 \\ 
	443 & Other peripheral vascular disease & 5.6 \\ 
	\MyIndent 4430 & Raynaud's syndrome & 0.3 \\ 
	\MyIndent 4438 & Peripheral angiopathy in diseases classified elsewhere & 0.3 \\ 
	\MyIndent 4439 & Peripheral vascular disease, unspecified & 5 \\ 
	444 & Arterial embolism and thrombosis & 0.2 \\ 
	\MyIndent 4442 & Arterial embolism and thrombosis of upper extremity & 0.1 \\ 
	\MyIndent 4448 & Embolism and thrombosis of iliac artery & 0.1 \\ 
	447 & Other disorders of arteries and arterioles & 0.5 \\ 
	\MyIndent 4471 & Stricture of artery & 0.4 \\ 
	\MyIndent 4477 & Aortic ectasia, unspecified site & 0.1 \\ 
	458 & Hypotension & 1.7 \\ 
	\MyIndent 4580 & Orthostatic hypotension & 0.2 \\ 
	\MyIndent 4582 & Hypotension of hemodialysis & 0.1 \\ 
	\MyIndent 4588 & Other specified hypotension & 0.6 \\ 
	\MyIndent 4589 & Hypotension, unspecified & 0.7 \\ 
	493 & Asthma & 4.9 \\ 
	\MyIndent 4932 & Chronic obstructive asthma, unspecified & 1.4 \\ 
	\MyIndent 4939 & Asthma,unspecified type, unspecified & 3.6 \\ 
	518 & Other diseases of lung & 3 \\ 
	\MyIndent 5180 & Pulmonary collapse & 0.4 \\ 
	\MyIndent 5184 & Acute edema of lung, unspecified & 0.1 \\ 
	\MyIndent 5188 & Acute respiratory failure & 2.5 \\ 
	530 & Diseases of esophagus & 17.3 \\ 
	\MyIndent 5301 & Esophagitis, unspecified & 0.3 \\ 
	\MyIndent 5308 & Esophageal reflux & 17.1 \\ 
	536 & Disorders of function of stomach & 0.3 \\ 
	\MyIndent 5363 & Gastroparesis & 0.1 \\ 
	\MyIndent 5368 & Dyspepsia and other specified disorders of function of stomach & 0.2 \\ 
	564 & Functional digestive disorders, not elsewhere classified & 1.3 \\ 
	\MyIndent 5640 & Constipation, unspecified & 0.8 \\ 
	\MyIndent 5641 & Irritable bowel syndrome & 0.6 \\ 
	571 & Chronic liver disease and cirrhosis & 0.6 \\ 
	\MyIndent 5715 & Cirrhosis of liver w/o mention of alcohol & 0.2 \\ 
	\MyIndent 5718 & Other chronic nonalcoholic liver disease & 0.4 \\ 
	578 & Gastrointestinal hemorrhage & 0.4 \\ 
	\MyIndent 5781 & Blood in stool & 0.1 \\ 
	\MyIndent 5789 & Hemorrhage of gastrointestinal tract, unspecified & 0.2 \\ 
	584 & Acute renal failure & 2.8 \\ 
	\MyIndent 5845 & Acute kidney failure with lesion of tubular necrosis & 0.3 \\ 
	\MyIndent 5849 & Acute kidney failure, unspecified & 2.5 \\ 
	585 & Chronic renal failure & 9.9 \\ 
	\MyIndent 5852 & Chronic kidney disease, Stage II (mild) & 0.4 \\ 
	\MyIndent 5853 & Chronic kidney disease, Stage III (moderate) & 1.9 \\ 
	\MyIndent 5854 & Chronic kidney disease, Stage IV (severe) & 0.6 \\ 
	\MyIndent 5855 & Chronic kidney disease, Stage V & 0.2 \\ 
	\MyIndent 5856 & End stage renal disease & 1.8 \\ 
	\MyIndent 5859 & Chronic kidney disease, unspecified & 5.1 \\ 
	599 & Other disorders of urethra and urinary tract & 1.9 \\ 
	\MyIndent 5990 & Urinary tract infection, site not specified & 1.5 \\ 
	\MyIndent 5996 & Urinary obstruction, unspecified & 0.1 \\ 
	\MyIndent 5997 & Hematuria, unspecified & 0.3 \\ 
	696 & Psoriasis and similar disorders & 0.8 \\ 
	\MyIndent 6960 & Psoriatic arthropathy & 0.2 \\ 
	\MyIndent 6961 & Other psoriasis & 0.6 \\ 
	707 & Chronic ulcer of skin & 0.8 \\ 
	\MyIndent 7070 & Pressure ulcer, unspecified site & 0.3 \\ 
	\MyIndent 7071 & Ulcer of lower limb, unspecified & 0.5 \\ 
	\MyIndent 7072 & Pressure ulcer, unspecified stage & 0.3 \\ 
	710 & Diffuse diseases of connective tissue & 0.4 \\ 
	\MyIndent 7100 & Systemic lupus erythematosus & 0.2 \\ 
	\MyIndent 7101 & Systemic sclerosis & 0.2 \\ 
	715 & Osteoarthrosis and allied disorders & 4 \\ 
	\MyIndent 7153 & Osteoarthrosis, localized & 0.8 \\ 
	\MyIndent 7159 & Osteoarthrosis, unspecified & 3.2 \\ 
	721 & Spondylosis and allied disorders & 0.7 \\ 
	\MyIndent 7210 & Cervical spondylosis w/o myelopathy & 0.2 \\ 
	\MyIndent 7213 & Lumbosacral spondylosis w/o myelopathy & 0.3 \\ 
	\MyIndent 7219 & Spondylosis of unspecified site, w/o mention of myelopathy & 0.2 \\ 
	722 & Intervertebral disc disorders & 0.2 \\ 
	\MyIndent 7221 & Displacement of lumbar intervertebral disc w/o myelopathy & 0.1 \\ 
	\MyIndent 7225 & Degeneration of thoracic or thoracolumbar intervertebral disc & 0.1 \\ 
	723 & Other disorders of cervical region & 0.3 \\ 
	\MyIndent 7231 & Cervicalgia & 0.2 \\ 
	\MyIndent 7234 & Brachial neuritis or radiculitis NOS & 0.1 \\ 
	724 & Other and unspecified disorders of back & 3.4 \\ 
	\MyIndent 7240 & Spinal stenosis, unspecified region & 0.9 \\ 
	\MyIndent 7242 & Lumbago & 1.1 \\ 
	\MyIndent 7243 & Sciatica & 0.2 \\ 
	\MyIndent 7245 & Backache, unspecified & 1.3 \\ 
	729 & Other disorders of soft tissues & 1.3 \\ 
	\MyIndent 7291 & Myalgia and myositis, unspecified & 0.8 \\ 
	\MyIndent 7295 & Pain in limb & 0.1 \\ 
	\MyIndent 7298 & Swelling of limb & 0.4 \\ 
	733 & Other disorders of bone and cartilage & 2.1 \\ 
	\MyIndent 7330 & Osteoporosis, unspecified & 1.7 \\ 
	\MyIndent 7339 & Disorder of bone and cartilage, unspecified & 0.4 \\ 
	780 & General symptoms & 3.5 \\ 
	\MyIndent 7800 & Coma & 0.2 \\ 
	\MyIndent 7802 & Syncope and collapse & 0.7 \\ 
	\MyIndent 7803 & Febrile convulsions (simple), unspecified & 0.2 \\ 
	\MyIndent 7804 & Dizziness and giddiness & 0.4 \\ 
	\MyIndent 7805 & Sleep disturbance, unspecified & 1.4 \\ 
	\MyIndent 7806 & Fever, unspecified & 0.3 \\ 
	\MyIndent 7807 & Chronic fatigue syndrome & 0.2 \\ 
	\MyIndent 7809 & Fussy infant (baby) & 0.3 \\ 
	785 & Symptoms involving cardiovascular system & 2.4 \\ 
	\MyIndent 7850 & Tachycardia, unspecified & 0.2 \\ 
	\MyIndent 7852 & Undiagnosed cardiac murmurs & 0.3 \\ 
	\MyIndent 7855 & Shock, unspecified & 2 \\ 
	786 & Symptoms involving chest or respiratory system  & 1.5 \\ 
	\MyIndent 7860 & Respiratory abnormality, unspecified & 0.6 \\ 
	\MyIndent 7862 & Cough & 0.1 \\ 
	\MyIndent 7863 & Hemoptysis & 0.1 \\ 
	\MyIndent 7865 & Chest pain, unspecified & 0.7 \\ 
	787 & Symptoms involving digestive system & 0.8 \\ 
	\MyIndent 7870 & Nausea with vomiting & 0.2 \\ 
	\MyIndent 7872 & Dysphagia, unspecified & 0.3 \\ 
	\MyIndent 7879 & Diarrhea & 0.3 \\ 
	788 & Symptoms involving urinary system & 0.7 \\ 
	\MyIndent 7882 & Retention of urine, unspecified & 0.3 \\ 
	\MyIndent 7883 & Urinary incontinence, unspecified & 0.3 \\ 
	\MyIndent 7884 & Urinary frequency & 0.1 \\ 
	790 & Nonspecific findings on examination of blood & 2.4 \\ 
	\MyIndent 7902 & Impaired fasting glucose & 1.8 \\ 
	\MyIndent 7904 & Elevation of levels of transaminase or LDH & 0.2 \\ 
	\MyIndent 7906 & Other abnormal blood chemistry & 0.1 \\ 
	\MyIndent 7909 & Abnormal arterial blood gases & 0.3 \\ 
	794 & Nonspecific abnormal results of function studies & 1.4 \\ 
	\MyIndent 7943 & Abnormal cardiovascular function study, unspecified & 1.2 \\ 
	\MyIndent 7948 & Nonspecific abnormal results of function study of liver & 0.2 \\ 
	996 & Complications peculiar to certain specified procedures & 5 \\ 
	\MyIndent 9967 & Other complications due to unspecified device, implant, and graft & 4.8 \\ 
	\MyIndent 9968 & Complications of transplanted organ, unspecified & 0.2 \\ 
	E87 &  & 0.5 \\ 
	\MyIndent E878 & Transplant of whole organ causing abnormal reaction or complication & 0.4 \\ 
	\MyIndent E879 & Cardiac catheterization as the cause of reaction or complication & 0.1 \\ 
	\hline
	
	\MyIndent 2780 & Obesity, unspecified & 12.3 \\ 
	\MyIndent 4039 & Hypertensive chronic kidney disease & 9.1 \\ 
	\MyIndent 4139 & Other and unspecified angina pectoris & 7.9 \\ 
	\MyIndent 496 & Chronic airway obstruction, unspecified & 7.3 \\ 
	\MyIndent 3272 & Organic sleep apnea, unspecified & 4.9 \\ 
	\MyIndent 6000 & Hypertrophy of prostate without urinary obstruction & 4.9 \\ 
	\MyIndent 311 & Depressive disorder, unspecified & 4.8 \\ 
	\MyIndent 4168 & Other chronic pulmonary heart diseases & 1.9 \\ 
	\MyIndent 3572 & Polyneuropathy in diabetes & 1.8 \\ 
	\MyIndent 7140 & Rheumatoid arthritis & 1.5 \\ 
	\MyIndent 7169 & Arthropathy, unspecified, site unspecified & 1.5 \\ 
	\MyIndent 3659 & Unspecified glaucoma & 1.4 \\ 
	\MyIndent 486 & Pneumonia, organism unspecified & 1.3 \\ 
	\MyIndent 5621 & Diverticulosis of colon & 1.2 \\ 
	\MyIndent 3970 & Diseases of tricuspid valve & 1.2 \\ 
	\MyIndent 4912 & Obstructive chronic bronchitis without exacerbation & 1.1 \\ 
	\MyIndent 2886 & Leukocytosis, unspecified & 1.1 \\ 
	\MyIndent 3459 & Epilepsy & 1.1 \\ 
	\MyIndent 5533 & Diaphragmatic hernia & 0.9 \\ 
	\MyIndent 3569 & Unspecified peripheral neuropathy & 0.9 \\ 
	\MyIndent 3039 & Alcohol dependence & 0.9 \\ 
	\MyIndent V498 & Asymptomatic postmenopausal status & 0.8 \\ 
	\MyIndent 2689 & Vitamin D deficiency & 0.8 \\ 
	\MyIndent 3469 & Migraine & 0.8 \\ 
	\MyIndent 5838 & Nephritis and nephropathy & 0.8 \\ 
	\MyIndent 4598 & Venous insufficiency & 0.7 \\ 
	\MyIndent 5939 & Unspecified disorder of kidney and ureter & 0.7 \\ 
	\MyIndent 4928 & Other emphysema & 0.7 \\ 
	\MyIndent 7194 & Pain in joint & 0.6 \\ 
	\MyIndent 515 & Postinflammatory pulmonary fibrosis & 0.6 \\ 
	\MyIndent 3899 & Unspecified hearing loss & 0.6 \\ 
	\MyIndent 185 & Malignant neoplasm of prostate & 0.5 \\ 
	\MyIndent 2777 & Dysmetabolic syndrome & 0.5 \\ 
	\MyIndent 725  & Polymyalgia rheumatica & 0.5 \\ 
	\MyIndent 3098 & Posttraumatic stress disorder & 0.5 \\ 
	\MyIndent 6078 & Balanitis xerotica obliterans & 0.4 \\ 
	\MyIndent 5119 & Unspecified pleural effusion & 0.4 \\ 
	\MyIndent 3320 & Paralysis agitans & 0.4 \\ 
	\MyIndent 2662 & Other B-complex deficiencies & 0.4 \\ 
	\MyIndent 2898 & Primary hypercoagulable state & 0.4 \\ 
	\MyIndent 3694 & Legal blindness, as defined in U.S.A. & 0.4 \\ 
	\MyIndent 7990 & Asphyxia & 0.4 \\ 
	\MyIndent 9981 & Hemorrhage complicating a procedure & 0.4 \\ 
	\MyIndent 4029 & Hypertensive heart disease without heart failure & 0.4 \\ 
	\MyIndent 5569 & Ulcerative colitis & 0.4 \\ 
	\MyIndent 412 & Old myocardial infarction & 0.4 \\ 
	\MyIndent 5070 & Pneumonitis due to inhalation of food or vomitus & 0.4 \\ 
	\MyIndent 7840 & Headache & 0.3 \\ 
	\MyIndent 7823 & Edema & 0.3 \\ 
	\MyIndent 5559 & Regional enteritis of unspecified site & 0.3 \\ 
	\MyIndent V08 & Asymptomatic human immunodeficiency virus & 0.3 \\ 
	\MyIndent 3669 & Unspecified cataract & 0.3 \\ 
	\MyIndent 9971 & Cardiac complications, not elsewhere classified & 0.3 \\ 
	\MyIndent 5355 & Gastritis and gastroduodenitis & 0.3 \\ 
	\MyIndent 5888 & Secondary hyperparathyroidism & 0.3 \\ 
	\MyIndent V458 & Aortocoronary bypass status & 0.3 \\ 
	\MyIndent 2041 & Chronic lymphoid leukemia& 0.3 \\ 
	\MyIndent 4779 & Allergic rhinitis& 0.3 \\ 
	\MyIndent E849 & Home accidents & 0.3 \\ 
	\MyIndent 2028 & Other malignant lymphomas & 0.2 \\ 
	\MyIndent 2520 & Hyperparathyroidism & 0.2 \\ 
	\MyIndent 5965 & Hypertonicity of bladder & 0.2 \\ 
	\MyIndent 2387 & Essential thrombocythemia & 0.2 \\ 
	\MyIndent 3310 & Alzheimer's disease & 0.2 \\ 
	\MyIndent 4239 & Disease of pericardium & 0.2 \\ 
	\MyIndent 3140 & Attention deficit disorder without mention of hyperactivity & 0.2 \\ 
	\MyIndent 5339 & Peptic ulcer of unspecified site & 0.2 \\ 
	\MyIndent 6929 & Contact dermatitis and other eczema & 0.2 \\ 
	\MyIndent 5920 & Calculus of kidney & 0.2 \\ 
	\MyIndent 9959 & Systemic inflammatory response syndrome & 0.2 \\ 
	\MyIndent 135 & Sarcoidosis & 0.2 \\ 
	\MyIndent 2918 & Alcohol withdrawal & 0.2 \\ 
	\MyIndent 340 & Malignant neoplasm of right main bronchus & 0.2 \\ 
	\MyIndent 6826 & Cellulitis and abscess of leg, except foot & 0.2 \\ 
	\MyIndent 7890 & Abdominal pain & 0.2 \\ 
	\MyIndent 042 & Human immunodeficiency virus  & 0.2 \\ 
	\MyIndent 4049 & Hypertensive heart and chronic kidney disease, unspecified & 0.2 \\ 
	\MyIndent 4465 & Giant cell arteritis & 0.2 \\ 
	\MyIndent 7288 & Interstitial myositis & 0.2 \\ 
	\MyIndent 4824 & Pneumonia due to Staphylococcus, unspecified & 0.2 \\ 
	\MyIndent 2410 & Nontoxic uninodular goiter & 0.2 \\ 
	\MyIndent 0389 & Unspecified septicemia & 0.1 \\ 
	\MyIndent 2841 & Antineoplastic chemotherapy induced pancytopenia & 0.1 \\ 
	\MyIndent 2959 & Unspecified schizophrenia, unspecified & 0.1 \\ 
	\MyIndent 3040 & Opioid type dependence, unspecified & 0.1 \\ 
	\MyIndent 3540 & Carpal tunnel syndrome & 0.1 \\ 
	\MyIndent 570 & Acute necrosis of liver & 0.1 \\ 
	\MyIndent 0084 & Intestinal infection due to staphylococcus & 0.1 \\ 
	\MyIndent 1629 & Malignant neoplasm of bronchus and lung & 0.1 \\ 
	\MyIndent 1985 & Secondary malignant neoplasm of bone and marrow & 0.1 \\ 
	\MyIndent 2030 & Multiple myeloma & 0.1 \\ 
	\MyIndent 4739 & Unspecified sinusitis (chronic) & 0.1 \\ 
	\MyIndent 501 & Asbestosis & 0.1 \\ 
	\MyIndent 5589 & Noninfectious gastroenteritis and colitis & 0.1 \\ 
	\MyIndent 5771 & Chronic pancreatitis & 0.1 \\ 
	\MyIndent 5790 & Celiac disease & 0.1 \\ 
	\MyIndent 7832 & Loss of weight & 0.1 \\ 
	\MyIndent 1889 & Malignant neoplasm of bladder & 0.1 \\ 
	\MyIndent 2019 & Hodgkin's disease & 0.1 \\ 
	\MyIndent 2536 & Other disorders of neurohypophysis & 0.1 \\ 
	\MyIndent 2554 & Glucocorticoid deficiency & 0.1 \\ 
	\MyIndent 2713 & Intestinal disaccharidase deficiencies and malabsorption & 0.1 \\ 
	\MyIndent 2904 & Vascular dementia, uncomplicated & 0.1 \\ 
	\MyIndent 3860 & Meniere's disease & 0.1 \\ 
	\MyIndent 4059 & Unspecified renovascular hypertension & 0.1 \\ 
	\MyIndent 4151 & Iatrogenic pulmonary embolism and infarction & 0.1 \\ 
	\MyIndent 490 &  Bronchitis, acute/chronic unspecified & 0.1 \\ 
	\MyIndent 6953 & Rosacea & 0.1 \\ 
	\MyIndent 2113 & Benign neoplasm of colon & 0.1 \\ 
	\MyIndent 3429 & Hemiplegia affecting unspecified side & 0.1 \\ 
	\MyIndent 4370 & Cerebral atherosclerosis & 0.1 \\ 
	\MyIndent 5742 & Calculus of gallbladder & 0.1 \\ 
	\MyIndent 591 & Hydronephrosis & 0.1 \\ 
	\MyIndent 7367 & Deformity of ankle and foot, acquired & 0.1 \\ 
	\MyIndent 7812 & Abnormality of gait & 0.1 \\ 
	\MyIndent V125 & History of unspecified circulatory disease & 0.1 \\ 
	\MyIndent V158 & History of failed moderate sedation & 0.1 \\ 
	\hline
	
\end{longtable}

\end{document}